\documentclass[aps,prl,preprint]{revtex4}

\begin{document}
\input{psfig}

\title{Superresolution and Corrections to the Diffusion Approximation \\ 
in Optical Tomography}

\author{George Y. Panasyuk, Vadim A. Markel and John C. Schotland}

\affiliation{Departments of Bioengineering and Radiology \\ University of Pennsylvania, Philadelphia, PA 19104}

\date{\today}

\begin{abstract}
We demonstrate that the spatial resolution of images in optical tomography
is not limited to the fundamental length scale of one transport mean free
path. This result is facilitated by the introduction of novel corrections
to the standard integral equations of scattering theory within the 
diffusion approximation to the radiative transport equation.
\end{abstract}
 
\maketitle

There has been considerable recent interest in the development of
optical methods for tomographic imaging~\cite{Gibson_2005}. The
physical problem that is considered is to recover the optical
properties of the interior of an inhomogeneous medium from
measurements taken on its surface. The starting point for the
mathematical formulation of this inverse scattering problem (ISP) is a
model for the propagation of light, typically taken to be the
diffusion approximation (DA) to the radiative transport equation
(RTE). The DA is valid when the energy density of the optical field
varies slowly on the scale of the transport mean free path $\ell^*$.
The DA breaks down in optically thin layers, near boundary surfaces,
or near the source. One or more of these conditions are encountered in
biomedical applications such as imaging of small
animals~\cite{Graves_2003} or of functional activity in the brain.

Within the accuracy of the DA, reconstruction algorithms based on both
numerical~\cite{Arridge_1999} and analytic
solutions~\cite{Schotland_1997, Markel_2002, Markel_2004_1} to the ISP
have been described. Regardless of the method of inversion, the
spatial resolution of reconstructed images is expected to be limited
to $\ell^*$. This expectation is due to the intertwined effects of the
ill-posedness of the ISP~\cite{Markel_2002} and intrinsic inaccuracies
of the DA~\cite{Yoo_1990}. In this Letter, we introduce novel
corrections to the integral equations of scattering theory within the
DA. Using this result, we report the reconstruction of {\em
  superresolved} images whose spatial resolution is less than
$\ell^*$.

We begin by considering the propagation of multiply-scattered light in
an inhomogenous medium characterized by an absorption coefficient
$\mu_a({\bf r})$.  In what follows, we will neglect the contribution
of ballistic photons and consider only diffuse photons whose specific
intensity $I({\bf r},\hat{\bf s})$ at the point ${\bf r}$ in the
direction $\hat{\bf s}$ is taken to obey the time-independent RTE
\begin{eqnarray}
\label{RTE}
\hat{\bf s}\cdot\nabla I({\bf r},\hat{\bf s})  +(\mu_a + \mu_s)I({\bf r},\hat{\bf s}) -\mu_s\int d^2 s' A(\hat{\bf s},\hat{\bf s}') I({\bf r},\hat{\bf s}')  = S({\bf r},\hat{\bf s}) \ ,
\end{eqnarray}
where $\mu_s$ is the scattering coefficient, $A(\hat{\bf s},\hat{\bf
  s}')$ is the scattering kernel, and $S({\bf r},\hat{\bf s})$ is the
source.  The change in specific intensity due to spatial fluctuations
in $\mu_a({\bf r})$ can be obtained from the integral equation
\begin{equation}
\label{int_eq}
\phi({\bf r}_1,\hat{\bf s}_1;{\bf r}_2,\hat{\bf s}_2) = \int d^3r d^2s G({\bf r}_1,\hat{\bf s}_1;{\bf r},\hat{\bf s})G({\bf r},\hat{\bf s};{\bf r}_2,\hat{\bf s}_2)\delta\mu_a({\bf r}) \ .
\end{equation}
Here the data function $\phi({\bf r}_1,\hat{\bf s}_1;{\bf
  r}_2,\hat{\bf s}_2)$ is proportional, to lowest order in
$\delta\mu_a$, to the change in specific intensity relative to a
reference medium with absorption $\mu_a^0$, $G$ is the Green's
function for (\ref{RTE}) with $\mu_a=\mu_a^0$, $\delta\mu_a({\bf r}) =
\mu_a({\bf r})-\mu_a^0$, ${\bf r}_1,\hat{\bf s}_1$ and ${\bf
  r}_2,\hat{\bf s}_2$ denote the position and direction of a
unidirectional point source and detector, respectively.

We now show that the integral equation (\ref{int_eq}) may be used to
obtain corrections to the usual formulation of scattering theory
within the DA. To proceed, we note that, following
Ref.~\cite{Markel_2004_1}, the Green's function $G({\bf r},\hat{\bf
  s};{\bf r}',\hat{\bf s}')$ may be expanded in angular harmonics of
$\hat{\bf s}$ and $\hat{\bf s}'$:
\begin{equation}
\label{Green}
G({\bf r},\hat{\bf s};{\bf r}',\hat{\bf s}')={c\over{4\pi}}\left(1+\ell^*\hat{\bf s}\cdot\nabla_{{\bf r}}
\right)\left(1-\ell^*\hat{\bf s}'\cdot\nabla_{{\bf r}'}\right)G({\bf r},{\bf r}') \ ,
\end{equation}
where $\ell^* = 1/[\mu_a^0+ \mu_s']$ with $\mu_s'= (1-g)\mu_s$, $g$
being the anisotropy of the scattering kernel $A$. The Green's
function $G({\bf r},{\bf r}')$ satisfies the diffusion equation
$\left(-D_0\nabla^2 + \alpha_0 \right) G({\bf r},{\bf r}')=\delta({\bf
  r}-{\bf r}')$, where the diffusion coefficient $D_0=1/3c\ell^*$ and
$\alpha_0=c\mu_a^0$. In addition, the Green's function must satisfy
boundary conditions on the surface of the medium (or at infinity in
the case of free boundaries). In general we will consider boundary
conditions of the form $G({\bf r},{\bf r}') + \ell \hat{\bf n} \cdot
\nabla G({\bf r},{\bf r}') = 0$, where $\hat{\bf n}$ is the outward
unit normal to the surface bounding the medium and $\ell$ is the
extrapolation distance. Making use of (\ref{Green}) and performing the
angular integration over $\hat{\bf s}$ in (\ref{int_eq}) we obtain
\begin{eqnarray}
\label{corrections}
\phi({\bf r}_1,\hat{\bf s}_1;{\bf r}_2,\hat{\bf s}_2) = {c\over 4\pi}\Delta_{1}\Delta_{2}
\int d^3 r 
\Bigg[ G({\bf r}_1,{\bf r})G({\bf r},{\bf r}_2) 
 - {{\ell^*}^2\over 3}\nabla_{\bf r} G({\bf r}_1,{\bf r}) \cdot \nabla_{\bf r} G({\bf r},{\bf r}_2) \Bigg]
\delta\alpha({\bf r}) \ ,
\end{eqnarray}
where the differential operators $\Delta_k = 1 - (-1)^k\ell^*\hat{\bf
  s}_k\cdot\nabla_{{\bf r}_k}$ with $k=1,2$ and
$\delta\alpha=c\delta\mu_a$. Note that if the source and detector are
oriented in the inward and outward normal directions, respectively,
then (\ref{corrections}) becomes
\begin{eqnarray}
\label{boundary_corrections}
\nonumber
\phi \Big({\bf r}_1,-\hat{\bf n}({\bf r}_1);{\bf r}_2,\hat{\bf n}({\bf r}_2)\Big) = {c\over 4\pi}\left(1+{\ell^*\over \ell}\right)^2 \int d^3 r \Bigg[ G({\bf r}_1,{\bf r}) 
G({\bf r},{\bf r}_2)  \\
- {{\ell^*}^2\over 3} \nabla_{\bf r} G({\bf r}_1,{\bf r}) \cdot \nabla_{\bf r} G({\bf r},{\bf r}_2) \Bigg] 
\delta\alpha({\bf r}) \ ,
\end{eqnarray}
where we have used the boundary conditions on $G$ to evaluate the
action of the $\Delta_k$ operators. Eq.~(\ref{boundary_corrections})
is the main theoretical result of this Letter.  It may be viewed as
providing corrections to the DA since the first term on the right hand
side of (\ref{boundary_corrections}) corresponds to the standard DA in
an inhomogeneous absorbing medium. We note that the second term may be
interpreted as defining an effective diffusion coefficient $D({\bf
  r})=D_0 - ({\ell^*}^2/3) \delta\alpha({\bf r})$ since the expression
$\nabla_{\bf r} G({\bf r}_1,{\bf r}) \cdot \nabla_{\bf r} G({\bf
  r},{\bf r}_2)$ defines the diffusion kernel in a medium with an
inhomogeneous diffusion coefficient~\cite{Arridge_1999}.

For the remainder of this paper we will work in the planar measurement
geometry, often encountered in small-animal imaging. In this case,
(\ref{corrections}) becomes
\begin{equation}
\label{int_eq_planar}
\phi({\bm rho}_1,{\bm rho}_2)=\int d^3r K({\bm \rho}_1,{\bm \rho}_2;{\bf r})
\delta\alpha({\bf r}) \ ,
\end{equation}
where ${\bm \rho}_1$ denotes the transverse coordinates of a point
source in the plane $z=0$, ${\bm \rho}_2$ denotes the transverse
coordinates of a point detector in the plane $z=L$, and the dependence
of $\phi$ on $\hat{\bf s}_1$ and $\hat{\bf s}_2$ is not made explicit.
Evidently, from considerations of invariance of the kernel $K({\bm
  \rho}_1,{\bm \rho}_2;{\bf r})$ under translations of its transverse
arguments, it can be seen that $K$ may be expressed as the Fourier
integral
\begin{eqnarray}
\label{def_K}
\nonumber
K({\bm \rho}_1,{\bm \rho}_2;{\bf r})={1\over(2\pi)^4}\int d^2q_1 d^2q_2 \kappa({\bf q}_1,{\bf q}_2;z) 
\exp\left[i({\bf q}_1-{\bf q}_2)\cdot{\bm \rho} - i ({\bf q}_1\cdot{\bm \rho}_1 - {\bf q}_2\cdot{\bm \rho}_2)\right] \ ,
\end{eqnarray}
where ${\bf r}=({\bm \rho},z)$.  The function $\kappa$ may be obtained
from the plane-wave expansion of the diffusion Green's function
obeying appropriate boundary conditions. In the case of free
boundaries, it is readily seen that $\kappa$ is given by the
expression
\begin{eqnarray}
\label{kappa}
\nonumber
\kappa({\bf q}_1,{\bf q}_2;z)={c\over 16\pi D_0^2 Q({\bf q}_1)Q({\bf q}_2)}
\left[1 +  {{\ell^*}^2\over3} 
\left(Q({\bf q}_1)Q({\bf q}_2)-{\bf q}_1\cdot{\bf q}_2 \right)\right] \\
\times  \left[1 + \ell^*\left(Q({\bf q}_1)+Q({\bf q}_2)\right) + {\ell^*}^2
Q({\bf q}_1)Q({\bf q}_2)\right]
\exp\left[-Q({\bf q}_1)|z| - Q({\bf q}_2)|z-L|\right] \ ,
\end{eqnarray}
where $Q({\bf q})=(q^2 + \alpha_0/D_0)^{1/2}$ and we have assumed that $\hat{\bf s}_1=\hat{\bf s}_2=\hat{\bf z}$. 

Inversion of the integral equation (\ref{int_eq_planar}) may be
carried out by analytic methods. These methods have been shown to be
computationally efficient and may be applied to data sets consisting
of a very large number of
measurements~\cite{Markel_2002,Markel_2004_1}.  The approach taken is
to construct the singular value decomposition of the linear operator
$K$ in the proper Hilbert space setting and then use this result to
obtain the pseudoinverse solution to (\ref{int_eq_planar}). In this
manner, it is possible to account for the effects of sampling and
thereby obtain the best (in the sense of minimizing the appropriate
$L^2$ norm) bandlimited approximation to $\delta\alpha$. Here we use
this approach to simulate the reconstruction of a point absorber
located at a point ${\bf r}_0$ between the measurement planes with
$\delta\alpha({\bf r})=A\delta({\bf r}-{\bf r}_0)$ for constant $A$.
In this situation it is possible to calculate the data function $\phi$
within radiative transport theory, thus avoiding ``inverse crime.'' To
proceed, we require the Green's function $G({\bf r},\hat{\bf s};{\bf
  r}',\hat{\bf s}')$ for the RTE in a homogeneous infinite medium
which we obtain as described in Ref.~\cite{Markel_2004_2}. Note that
in this case, the angular integration over $\hat{\bf s}$ in
(\ref{int_eq}) may be carried out analytically.

The effects of corrections to the DA were studied in numerical
simulations following the methods of Ref.~\cite{Markel_2002}. The
simulations were performed for a medium with optical properties
similar to breast tissue in the near infrared~\cite{Peters_1990}. The
background absorption and reduced scattering coefficients were given
by $\mu_a^0 = 0.03 \ {\rm cm}^{-1}$ and $\mu_s'= 10 \ {\rm cm}^{-1}$.
The scattering kernel was taken to be of Henyey-Greenstein type with
$A(\hat{\bf s},\hat{\bf s}')=\sum_{\ell=0}^\infty
g^{\ell}P_\ell(\hat{\bf s}\cdot\hat{\bf s}')$ and $g=0.98$. This
choice of parameters corresponds to $\ell^*=1\ {\rm mm}$ and $D_0=0.8
\ {\rm cm^2 ns^{-1}}$. The separation between the measurement planes
$L$ was varied in order to explore the effects of the corrections. A
single point absorber was placed at the midpoint of the measurement
planes with ${\bf r}_0=(0,0,L/2)$ and $A= 1 \ {\rm cm}^3 \ {\rm
  ns}^{-1}$. The sources and detectors were located on a square
lattice with spacing $h$. The total number of source-detector pairs
$N$ was varied, along with $h$, as indicated below. To demonstrate the
stability of the reconstruction in the presence of noise, Gaussian
noise of zero mean was added to the data at the 1\% level, relative to
the average signal. Note that the level of regularization was chosen
to be the same for all reconstructions.

Reconstruction of $\delta\alpha({\bf r})$ for a point absorber defines
the point spread function (PSF) of the reconstruction algorithm. The
resolution $\Delta x$ is defined as the half width at half maximum of
the PSF. In Fig.~1(a) we consider the case of a thick layer with
$L=6.6 \ {\rm cm}$. The above parameters were chosen to be $h=0.83 \ 
{\rm mm}$ and $N=1.5\times 10^9$. PSFs with and without corrections
are shown. It may be seen that the effect of the corrections is
negligible in the case of a thick layer and the resolution $\Delta x =
3.5 \ell^*$. For the case of a layer of intermediate thickness with
$L=1.1 \ {\rm cm}$, as shown in Fig.~1(b), the corrections have a more
significant effect. In particular, with $h=0.28 \ {\rm mm}$ and
$N=1.2\times 10^{11}$, we found that $\Delta x= 0.9\ell^*$ for the
uncorrected reconstruction and $\Delta x= 0.7 \ell^*$ for the
corrected reconstruction. The corrections are most significant for the
thinnest layer considered in this study, achieving a factor of two
improvement in resolution when $L=0.55 \ {\rm cm}$. In this case, with
$h=0.14 \ {\rm mm}$ and $N=1.9\times 10^{12}$, we found that $\Delta x
= 0.4\ell^*$ for the uncorrected case and $\Delta x = 0.2 \ell^*$ for
the corrected case as shown in Fig.~1(c). Note that the number of
source-detector pairs $N\approx 10^9 - 10^{12}$ may be achieved in
modern non-contact optical tomography systems~\cite{Ripoll_2004}.

In conclusion, we have described a series of corrections to the usual
formulation of the DA in optical tomography. We have found that these
corrections give rise to superresolved images with resolution below
$\ell^*$. Several comments on these results are necessary. First, the
effects of corrections were demonstrated to be most significant in
optically thin layers. However, corrections to the DA may also be
expected to be important for thick layers when inhomogeneities in the
absorption are located near the surface. Second, the results of this
study were obtained without resorting to so-called inverse crime. That
is, forward scattering data was obtained from the full RTE under
conditions when the DA is known to break down. Third, the use of
analytic reconstruction algorithms was essential for handling the
extremely large data employed in this study. Finally, we note that
higher order corrections to the DA are also expected to be important
for the nonlinear ISP.

This research was supported by the NIH under the grant P41RR02305 and
by the AFOSR under the grant F41624-02-1-7001.

\newpage

\centerline{\psfig{file=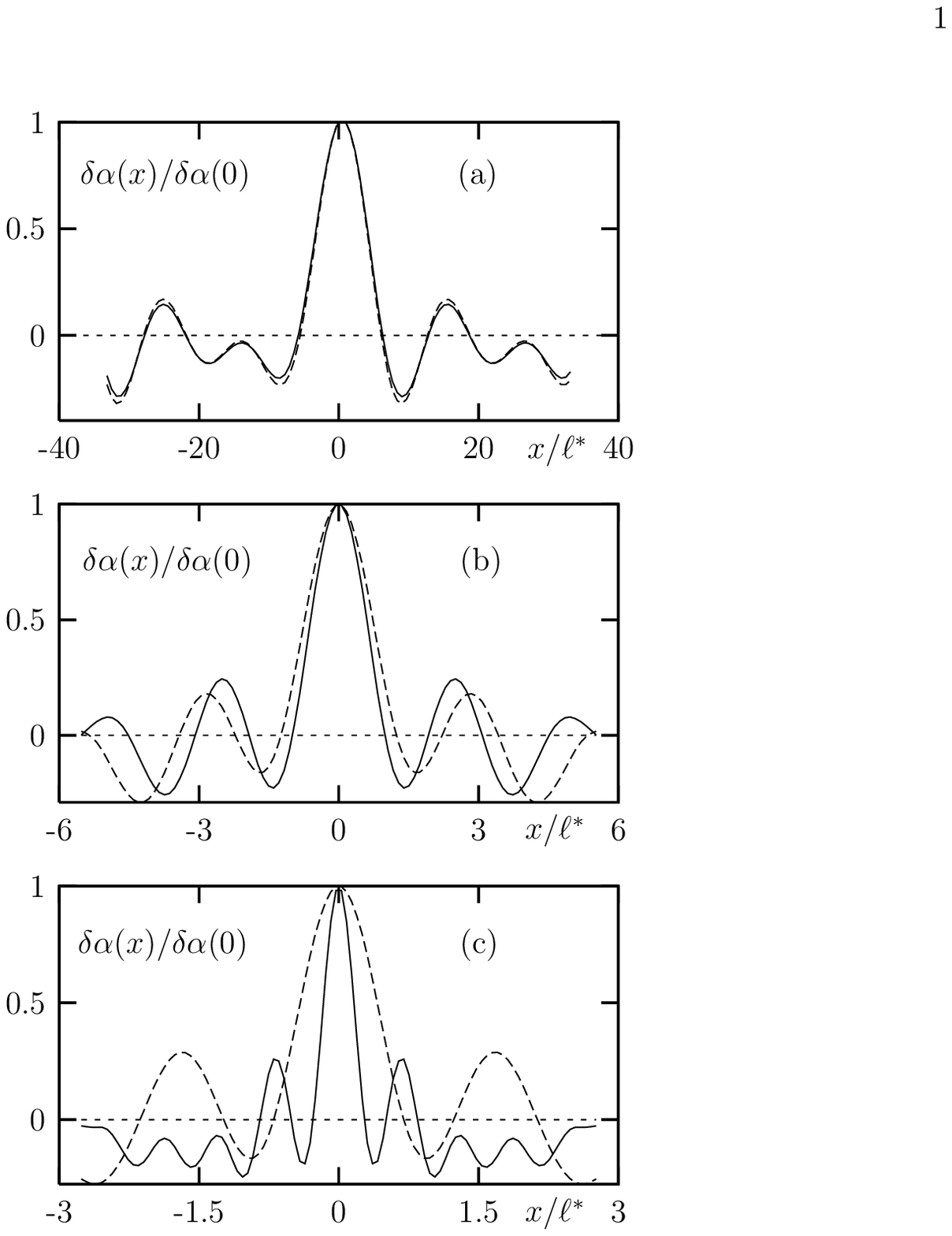}}
\vspace*{-10cm}
Fig. 1. Reconstructions of a point absorber for different thicknesses of the slab using the corrected (solid curve) and uncorrected (dashed curve) DA.

\end{document}